\documentclass[8.5pt,twoside,twocolumn]{article}
\usepackage[super,sort&compress,comma]{natbib} 
\usepackage{mhchem}
\usepackage{times,mathptmx}
\usepackage{sectsty}
\usepackage{balance} 
\usepackage{graphicx} %eps figures can be used instead
\usepackage{lastpage}
\usepackage[format=plain,justification=raggedright,singlelinecheck=false,font=small,labelfont=bf,labelsep=space]{caption} 
\usepackage{fancyhdr}
\newcommand{\beq}{ \begin{equation} }
\newcommand{\eeq}{ \end{equation} }
\newcommand{\beqs}{ \begin{eqnarray} }
\newcommand{\eeqs}{ \end{eqnarray} }
\newcommand{\f}[2]{\frac{#1}{#2}}

\newcommand{\upd}{\mathrm{d}}

\begin{document}

\thispagestyle{plain}
\fancypagestyle{plain}{
\fancyhead[L]{\includegraphics[height=8pt]{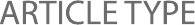}}
\fancyhead[C]{\hspace{-1cm}\includegraphics[height=20pt]{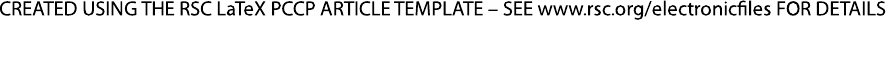}}
\fancyhead[R]{\includegraphics[height=10pt]{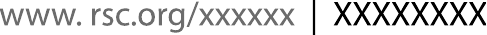}\vspace{-0.2cm}}
\renewcommand{\headrulewidth}{1pt}}
\renewcommand{\thefootnote}{\fnsymbol{footnote}}
\renewcommand\footnoterule{\vspace*{1pt}% 
\hrule width 3.4in height 0.4pt \vspace*{5pt}} 
\setcounter{secnumdepth}{5}

\makeatletter 
\def\subsubsection{\@startsection{subsubsection}{3}{10pt}{-1.25ex plus -1ex minus -.1ex}{0ex plus 0ex}{\normalsize\bf}} 
\def\paragraph{\@startsection{paragraph}{4}{10pt}{-1.25ex plus -1ex minus -.1ex}{0ex plus 0ex}{\normalsize\textit}} 
\renewcommand\@biblabel[1]{#1}            
\renewcommand\@makefntext[1]% 
{\noindent\makebox[0pt][r]{\@thefnmark\,}#1}
\makeatother 
\renewcommand{\figurename}{\small{Fig.}~}
\sectionfont{\large}
\subsectionfont{\normalsize} 

\fancyfoot{}
\fancyfoot[LO,RE]{\vspace{-7pt}\includegraphics[height=9pt]{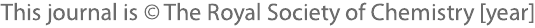}}
\fancyfoot[CO]{\vspace{-7.2pt}\hspace{12.2cm}\includegraphics{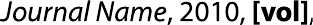}}
\fancyfoot[CE]{\vspace{-7.5pt}\hspace{-13.5cm}\includegraphics{headers/RF.pdf}}
\fancyfoot[RO]{\footnotesize{\sffamily{1--\pageref{LastPage} ~\textbar  \hspace{2pt}\thepage}}}
\fancyfoot[LE]{\footnotesize{\sffamily{\thepage~\textbar\hspace{3.45cm} 1--\pageref{LastPage}}}}
\fancyhead{}
\renewcommand{\headrulewidth}{1pt} 
\renewcommand{\footrulewidth}{1pt}
\setlength{\arrayrulewidth}{1pt}
\setlength{\columnsep}{6.5mm}
\setlength\bibsep{1pt}

\twocolumn[
  \begin{@twocolumnfalse}
\noindent\LARGE{\textbf{The collective motion of nematodes in a thin liquid layer\ddag}}
\vspace{0.6cm}

\noindent\large{\textbf{Sean Gart,$^{\ast}$ Dominic Vella,$^{\dag}$ and
Sunghwan Jung$^{\ast}$}} \vspace{0.5cm}
%Please note that \ast indicates the corresponding author(s) but no footnote text is required. 

\noindent\textit{\small{\textbf{Received Xth XXXXXXXXXX 20XX, Accepted Xth XXXXXXXXX 20XX\newline
First published on the web Xth XXXXXXXXXX 200X}}}

\noindent \textbf{\small{DOI: 10.1039/b000000x}}
\vspace{0.6cm}
%Please do not change this text.

\noindent \normalsize{Many organisms live in confined fluidic environments such as the thin liquid layers on the skin of host organisms or in partially-saturated soil. We investigate the collective behaviour of nematodes in a thin liquid layer, which was first observed by Gray and Lissmann, [\emph{J. Exp. Biol.} \textbf{41}, 135 (1964)]. We show experimentally that nematodes confined by a thin liquid film come into contact and only separate again after some intervention. We attribute this collective motion to an attractive force between them arising from the surface tension of the layer and show that for nearby nematodes this force is typically stronger than the force that may be exerted by the nematodes' muscles. We believe this to be the first demonstration of the ``Cheerios effect" acting on a living organism. However, we find that being grouped together does not significantly alter the body stroke and kinematic performance of the nematode: there are no statistically significant changes of the Strouhal number and the ratio of amplitude to wavelength when aggregated. This result implies that nematodes gain neither a mechanical advantage nor disadvantage by being grouped together; the capillary force merely draws and keeps them together.
}
\vspace{0.5cm}
 \end{@twocolumnfalse}
  ]

\section{Introduction}

%Please use \dag to cite the ESI in the main text of the article.
%If you article does not have ESI please remove the the \dag symbol from the title and the above footnotetext.

\footnotetext{\textit{$^{\ast}$~Department of Engineering Science and Mechanics, 
Virginia Polytechnic Institute and State University, Blacksburg, VA 24061, USA. 
Fax: 01 540-231-4574; Tel: 01 540-231-5146; E-mail: sunnyjsh@vt.edu}}
\footnotetext{\textit{$^{\dag}$~ITG, Department of Applied Mathematics and Theoretical Physics, University of Cambridge, Wilberforce Road, Cambridge, CB3 0WA, UK. }}
\footnotetext{\ddag~Electronic Supplementary Information (ESI) available: [Supplementary Videos 1 and 2]. See DOI: 10.1039/b000000x/}

%additional addresses can be cited as above using the lower-case letters, c, d, e... If all authors are from the same address, no letter is required

%%%%%%%%%%%%% Main Article %%%%%%%%%%%%%%%%%%%%%%%%%

Organisms need to move for a variety of reasons including to avoid predators, to find nutrients or to find a mate. This motion often takes place in fluid environments and hence may involve hydrodynamic interactions, which have been studied extensively in recent years. At the microscopic scale, hydrodynamic interactions between organisms can result in the generation of large-scale circulations \cite{PK1992,BZFPBS2009}, a highly concentrated layer in a shear field \cite{DKS2009} or the non-Brownian motion of a test particle \cite{WL2000}. This collective behaviour can also enhance mixing and transport of chemical nutrients \cite{TCDWKG2005,SAKG2007,KB2004}.
%; preferred swimming configurations have been observed during swarming motion \cite{BZFPBS2009}. 
In the majority of recent studies, collective movements take place in the bulk of fluids and hence owe their origin to bulk hydrodynamic interactions. 
Here we report the temporary aggregation of nematodes caused by the presence of a liquid--gas interface. 
Ostensibly, this is similar to the dancing of \emph{Volvox} (algal colonies) at the surface of water~\cite{Drescher2009}, but the physics is very different.

Nematodes locomote by creating undulatory movement similar to other invertebrates. The propagating body-wave is generated by contracting and relaxing muscles under the control of the neuromuscular system. 
The maximum force that nematodes are capable of generating has been measured in micro-fabricated devices, and shown to be on the order of 1 $\mu$N \cite{DNKKCPGP2009}. This limiting force suggests that the application of a force with magnitude greater than 1 $\mu$N may change or constrain the organism's behaviour.
%Mechano-sensors in nematodes typically respond to mechanical stimuli of more than 10 $\mu$N. % and change their behavior as a result \cite{Wormbook}. 
%***Does this mean that they should be changing their behavior because of the type of forces that we are dealing with here?*** {\bf I just mention it here to give an estimate of forces whether worms can feel. It does not mean they will change their behavior since we are not sure whether this capillary force gives a negative or positive response. . So, I delete the later part. }

%Since some nematodes live in a thin layer of liquid it is natural to suppose that their locomotion may be affected by surface tension. 
%We might also suppose that the thickness of a thin fluid film directly affects a nematode's undulating movement \cite{Wallace1968}. 
%It has been observed previously that in close proximity to other nematodes, they tend to aggregate and move together in a co--ordinated way \cite{GL1964}. 
%Qualitatively, such collective behavior has been found in many types of  nematodes such as the hookworm larvae and parasitic nematodes \cite{Croll1970b}. However, no explanation for this behaviour has previously been offered.

\begin{figure*}[ht] %% Exp_scheme 
    \centering
        \includegraphics[width=\textwidth]{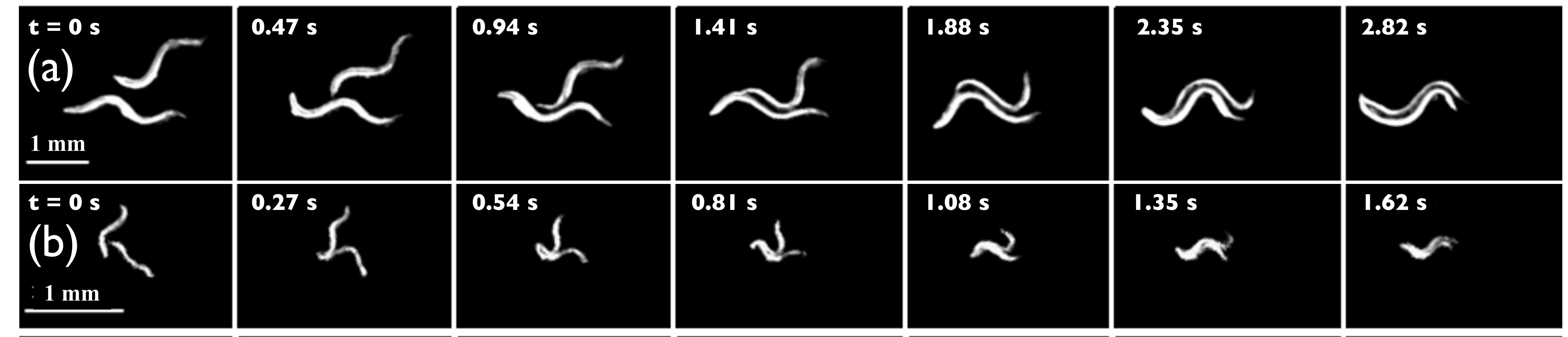}
\caption{Two nematodes ({\it Panagrellus redivivus}) merging: (a) two long nematodes and (b) two small nematodes. Typically, it takes about 1 sec for them to merge completely from the initial point of touching. }
    \label{Fig_nematodes}
\end{figure*}

It is often observed that nematodes migrate up the side of their culture jars and move into a thin liquid film region. When this occurs they are also often observed to aggregate together \cite{GL1964}. Qualitatively, such collective behaviour has been found in many types of  nematodes (such as the hookworm larvae and parasitic nematodes \cite{Croll1970b}) and has been shown \emph{not} to originate in chemical interactions \cite{Croll1970}. In this article we shall suggest that this aggregation is caused by lateral capillary forces --- similar to the ``the Cheerios effect" \cite{VM2005} observed with floating objects.

Capillary forces arise from the interfacial tension of fluid phase boundaries, and act along the interfaces. They can, therefore, be important for small organisms living close to fluid interfaces; they allow, for example, the pond skater to walk on the surface of water \cite{BH2006}.  When a liquid film is thinner than the diameter of a nematode body the nematode experiences both vertical and lateral forces because of the surface tension of the interface. The vertical component of surface tension pulls its body down against the substrate, lowering the lateral slipping during locomotion and allowing them to crawl on a surface. The horizontal component of surface tension, meanwhile, will vanish when the nematode is alone but may, in principle, lead to an interaction with a nearby nematode.

In this paper, we shall investigate the collective behaviour of nematodes in a thin liquid layer. We characterize the locomotory performance of nematodes by measuring the geometric parameters of their body stroke and forward mÄoving velocity. We compare this locomotory performance for the cases of aggregated and individual nematodes. We then present a simplified theoretical model to estimate the typical lateral forces due to surface tension and compare it with the typical forces generated by nematodes during locomotion. 
Our study rationalizes the physical basis of the aggregation but demonstrates ultimately that it changes little about the locomotion stroke.

\section{Experimental methods}

\subsection{Nematodes crawling on a thin fluid}
{\it Panagrellus redivivus} was used as the model organism in our experiments. {\it P. redivivus}, often called the micro-worm, is a free living, bacteriophagous nematode. 
It can be found in soil, vegetation, and dung \cite{Gray1983}, demonstrating its ability to move in various fluidic conditions.  
Cultures of {\it P. redivivus} were maintained at a temperature of $18~^\circ\mathrm{C}$ in a potato medium. % from the Carolina Biological Supply. 
{\it P. redivivus} grows to around 1 mm in length $L$ and 50 $\mu$m in diameter $2r_c$. 
\begin{figure}[bt] %% Exp_scheme 
    \centering
        \includegraphics[width=.5\textwidth]{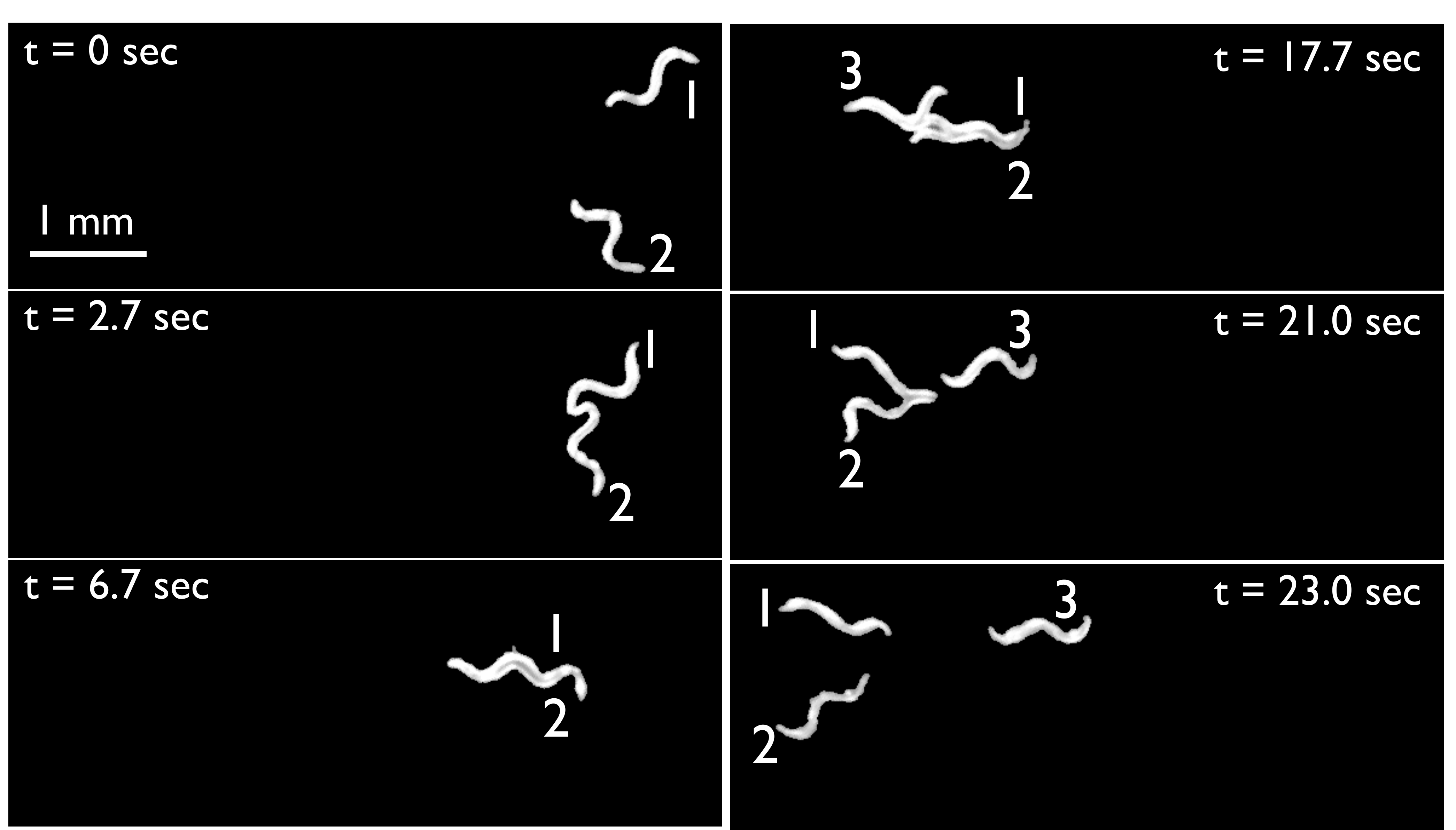}
\caption{Various stages of aggregating, joint, and separating nematodes. Two nematodes are aggregated at $t = 6.7$ sec, and stay together for more than 10 sec. However, they are hit by a third nematode at $t = 17.7$ sec ultimately leading to the separation of all three nematodes. (Supplementary Video 1$\ddag$) }
    \label{Fig_MSS}
\end{figure}

Nematode cultures were placed in sealed glass chambers. Over several hours some nematodes migrate up the side of the glass. We characterize the motion of nematodes a few centimetres above the bulk of the culture medium. At this height, only a thin liquid layer remains on the side glass wall, which is quite different from the slurry bulk culture below. Videos of nematodes are taken using a Sony Handycam with a macro lens and Virtualdub software was used to format video. Data are taken from the same chamber with the same ambient air temperature to avoid any thermotatic behaviours. 
The MATLAB image processing toolbox is used to analyze footage and process data.  
Peak-to-peak amplitude $A$ and wavelength $\lambda$ of the nematodes are measured at each stage of the aggregation process. 
Nematode speed $V$ is recorded by tracking the centroid of the body typically over a few periods of undulation. 
The frequency $f$ of nematode undulation and the length $L$ of each nematode are also measured.

Figure \ref{Fig_nematodes} shows two nematodes coming together as they crawl in proximity to each other. 
Typically, their heads touch and their forward motions draw the rest of the bodies together. 
Once they are grouped, the capillary force makes it difficult for the nematodes to separate themselves. 
We observe that nematodes stay together from as little as a few seconds to as long as a minute depending on whether they are hit by others (see Fig. \ref{Fig_MSS}). 

We often observed events in which two nematodes of different sizes aggregate as in Fig. \ref{Fig_nematodes}(b). 
Before two nematodes merge, the individuals have different body strokes and speed. Typically, smaller nematodes have shorter wavelength and amplitude and lower speed. However, after they aggregate they form a single body form --- the smaller nematode adopting the body form of the larger nematode.

% \begin{figure}[bt] %% Exp_scheme 
%    \centering
%        \includegraphics[width=.5\textwidth]{Fig_schematics}
%\caption{Images of two nematodes (a) aggregating (b) joint, and (c) separating. Physical quantities are measured at different stages. }
%    \label{Fig_schematics}
%\end{figure}

%Figure \ref{Fig_schematics} shows various measurements while two nematodes are aggregating, joining, and separating. 
%In cases of nematode separation or aggregation, maximum separation distances ($L_a, L_s$) and opening angles ($\theta_a, \theta_s$) were measured at the initial touching or final division of two nematodes. 

\begin{figure}[ht] %% Exp_scheme 
    \centering
        \includegraphics[width=.35\textwidth]{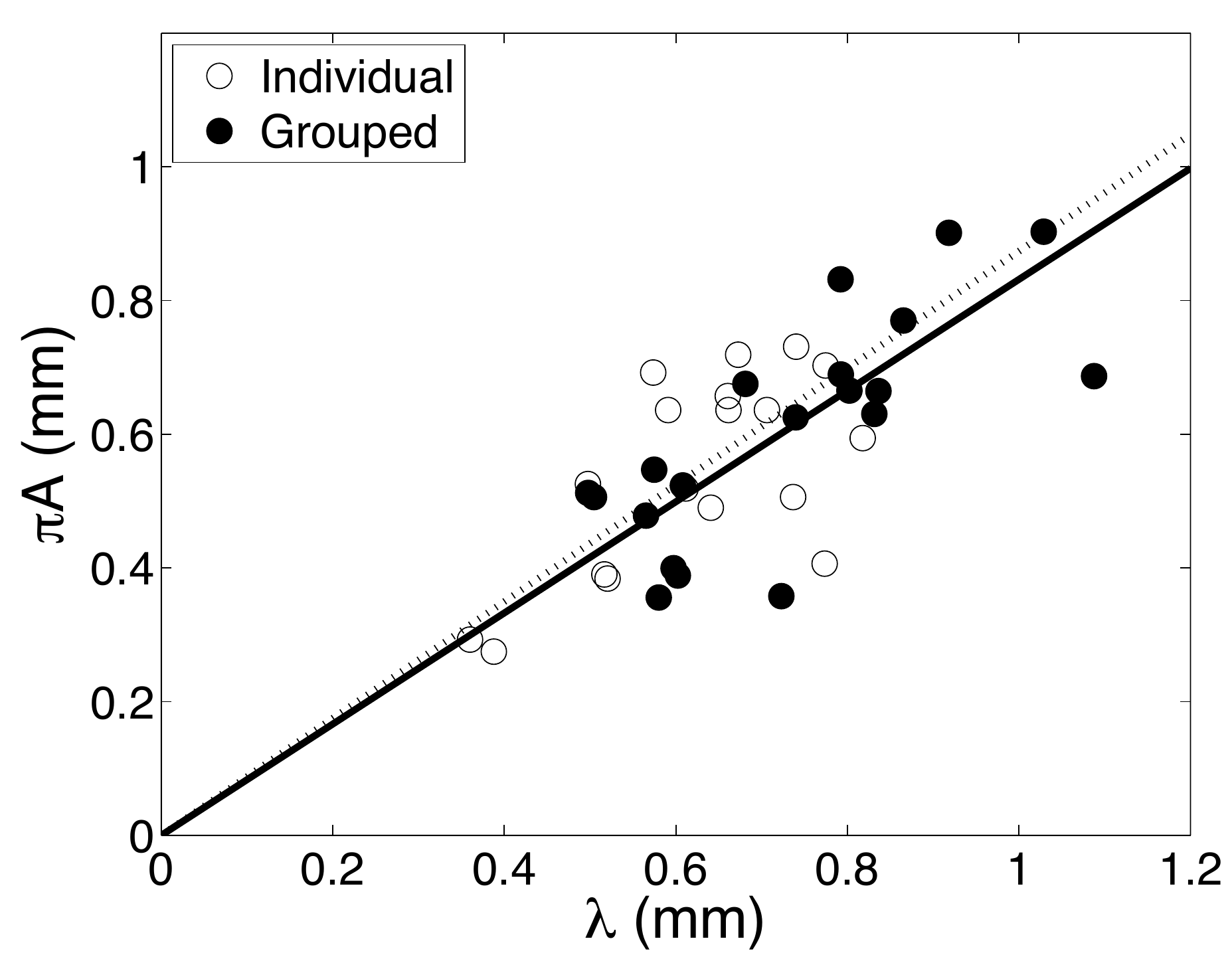}%{dv_ratio} \\
        \\ \includegraphics[width=.35\textwidth]{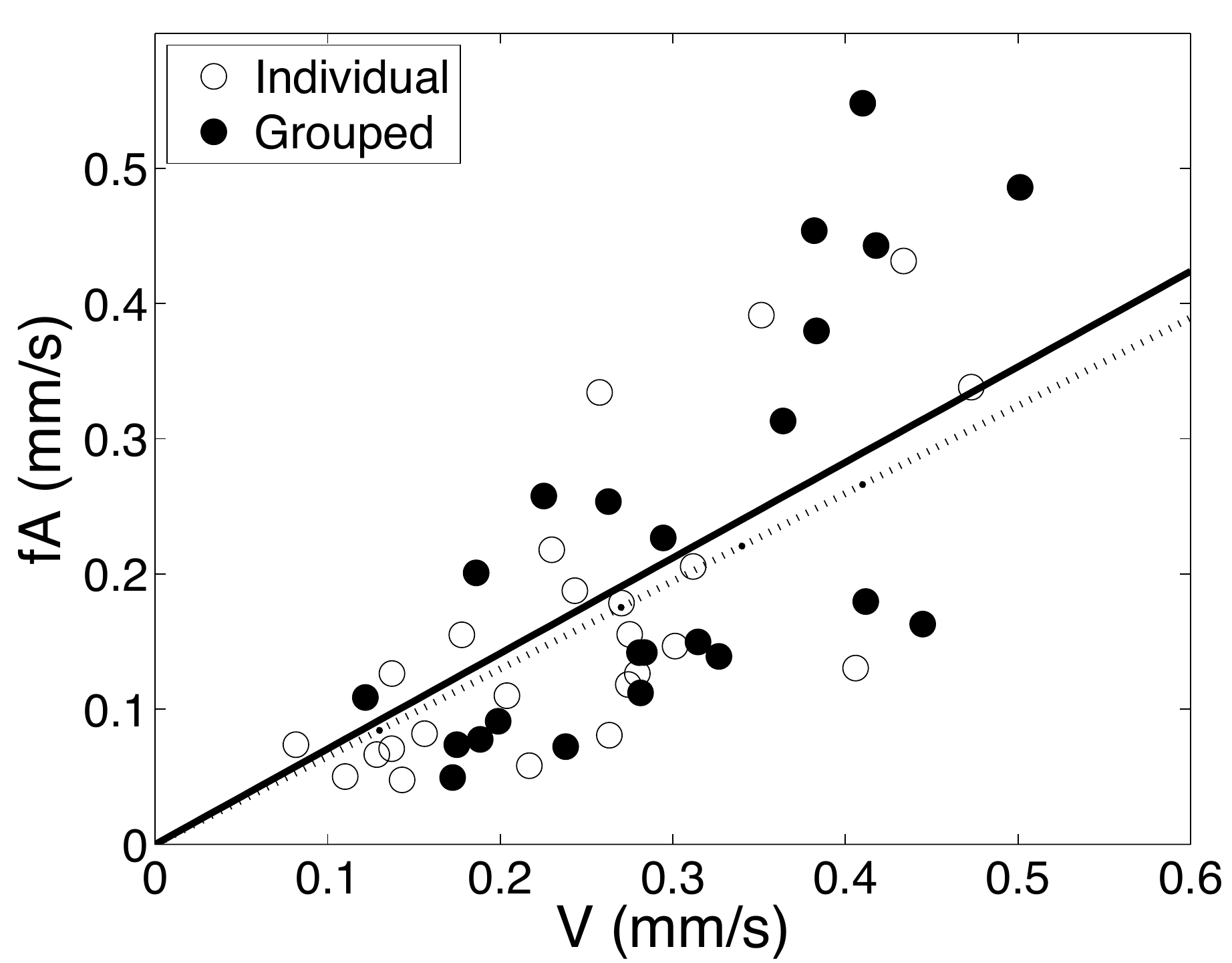}%{dv_Strouhal}
\caption{(a) Amplitude versus wavelength, and (b)  $fA$ versus forward velocity, $V$, for individual and grouped nematodes. In each case, the lines correspond to the mean ratios for individual (dashed line) and grouped (solid line) nematodes.}
    \label{Fig_faV}
\end{figure}

To characterize the aggregation behaviour, the  peak-to-peak amplitude over wavelength and non-dimensional Strouhal number (the ratio of lateral to forward velocities) were calculated for individual and paired nematodes. 
The ratio of waveform amplitude to wavelength is a primary indicator of changing body stroke in swimming organisms \cite{Holwill1977} while the Strouhal number gives an indication of the efficiency of the locomotion stroke \cite{TNT2003}. Data are taken from 24 individual nematodes, and 23 cases of grouped nematodes and the $95\%$ confidence intervals for mean ratio and mean Strouhal number are constructed for each condition (individual or grouped). The $95\%$ confidence interval for the mean ratio of amplitude to wavelength was $(0.79,0.95)$ amongst individuals whilst amongst grouped nematodes the corresponding confidence interval was $(0.76,0.90)$. Since these confidence intervals overlap, we conclude that there is no significant difference between the mean ratio of stroke amplitude to wavelength. However, we note that these values are higher than the value reported in crawling nematodes on a soft agar gel (0.63$\pm$0.03) \cite{KCSMCS2006}. The $95\%$ confidence interval of mean Strouhal number for individuals was $(0.54,0.76)$ while that for grouped nematodes was $(0.57,0.84)$. Since these confidence intervals overlap, we again conclude that there is no statistically significant difference between the mean Strouhal number for grouped and individual nematodes. In conclusion, the features of their locomotion are not changed by being aggregated. Figure \ref{Fig_faV} demonstrates  this similarity in the locmotion characteristics for individual  and grouped nematodes graphically.
%Individual nematodes had a marginally wider variety of movement characteristics than grouped nematodes as indicated by the lower error percentage of amplitude over wavelength values for grouped nematodes.  

\begin{figure}[ht] %% Exp_scheme 
    \centering
	\includegraphics[width=.4\textwidth]{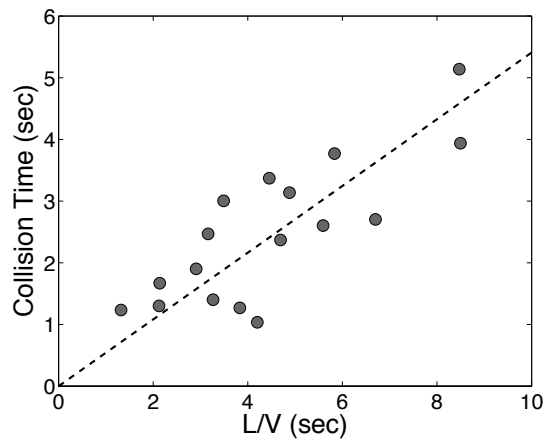}\\
\caption{Collision time vs. crawling time. Collision time is measured as the time interval from point-wise touch to grouped two-body  
and crawling time is defined as the ratio of body length to crawling velocity. }
    \label{Fig_Time}
\end{figure}

To characterize the dominant mechanism for nematode aggregation, we consider two possible time-scales associated with the aggregation. 
One is the capillary time-scale accounting for the surface tension force that draws them together, defined as $T_{capillary} = 6 \pi \mu r_c /\gamma$ where $\mu$ is the fluid viscosity ($4.0$ cP for a filtered potato medium) and $\gamma$ is the surface tension coefficient (note that this holds even for zipping bodies \cite{vella04}). 
The other is the crawling time defined as $T_{crawl} = L/V$. 

Experimentally, we define the collision time as the time interval from the initial contact to the grouped two-body. 
The grouped two-body is referred as a configuration when the two nematodes are touching along their body length but do not necessarily have their heads and tails in the same phase.
For example, the second column in Fig. \ref{Fig_nematodes}(a) is when two nematodes make the initial contact and the sixth column represents the grouped two-body. Hence, in Fig. \ref{Fig_nematodes}(a) the collision time is 1.88 sec. A comparison between crawling and observed collision time scales is shown in Fig. \ref{Fig_Time}. 
% The observed separation time scale was 0.230 $\pm$ 2.19 sec in the same order of the collision-driven time scale. 
We observe that the capillary time scale is expected to be on the order of $10^{-5}$ sec and hence is significantly smaller than the measured collision time scales. Furthermore, the scaling for $T_{capillary}$ does not explain the observed relationship between size, speed and collision time. However, the crawling time depends on the body size and the crawling velocity, and is on the order of the measured collision times. 
From this, it seems clear that the initial aggregation is driven by crawling, rather than by capillary forces.
We therefore hypothesize that aggregation initially occurs because of collisions but that, once aggregated, the capillary force keeps the two nematodes in one body form.

\section{Model}

%To estimate the relative magnitude of various forces, a few non-dimensional numbers are considered. Reynolds number is defined as a ratio of inertia to viscous effects,  Re = $ \rho V L/ \mu \sim  O(0.1)$. 
%The viscous stress is larger than the inertia force of the body.
%Bond number is defined as a ratio of gravity to surface tension,  Bo = $ \rho g L^2/\gamma \sim O(0.1)$. 
%The gravity is small compared to surface tension. Moreover, our experiment sets up the fluid interface parallel to gravity, in which the gravity effect is not needed to describe the interfacial surface.
%The Weber number is defined as a ratio of inertia to surface tension,  We = $\rho V^2 L/\gamma \sim O(10^{-6})$. 
%These non-dimensional numbers imply that the inertia and gravitational effects are negligible, but the surface tension dominates.
%We can infer two stages where the surface tension affects the biolocomotion; 
%One is when two nematodes are about to merge, like zipping two flexible bodies \cite{vella04}. 
%However, this hypothesis has been proven to be false by measuring collision time-scales.
%The other is the capillary force that holds bodies tightly once they are joined.

We now seek to rationalize the observation that nematodes seem to be bound together after having come close to one another. We begin by considering the relative importance of the various physical forces in the system. Firstly, we note that the Bond number, $Bo=\rho g L^2/\gamma \sim O(0.1)$: thus surface tension dominates gravity for nematodes. In what follows, therefore, we shall neglect gravity entirely. We also note that the Reynolds number, which measures the relative importance of inertia to viscous effects,  $Re=\rho V L/ \mu \sim  O(10^{-3})$ and so liquid inertia is entirely negligible here. 
The insignificance of inertia means that velocities are proportional to forces and it is enough to calculate the typical interaction force between two nearby nematodes. In principle, it would then be possible to calculate the perturbation caused to their motion by this force, along the lines of capillary zippering \cite{vella04}, though we have already seen that this perturbation must be small. 

\begin{figure}[ht] %% Exp_scheme 
    \centering
        \includegraphics[width=.4\textwidth]{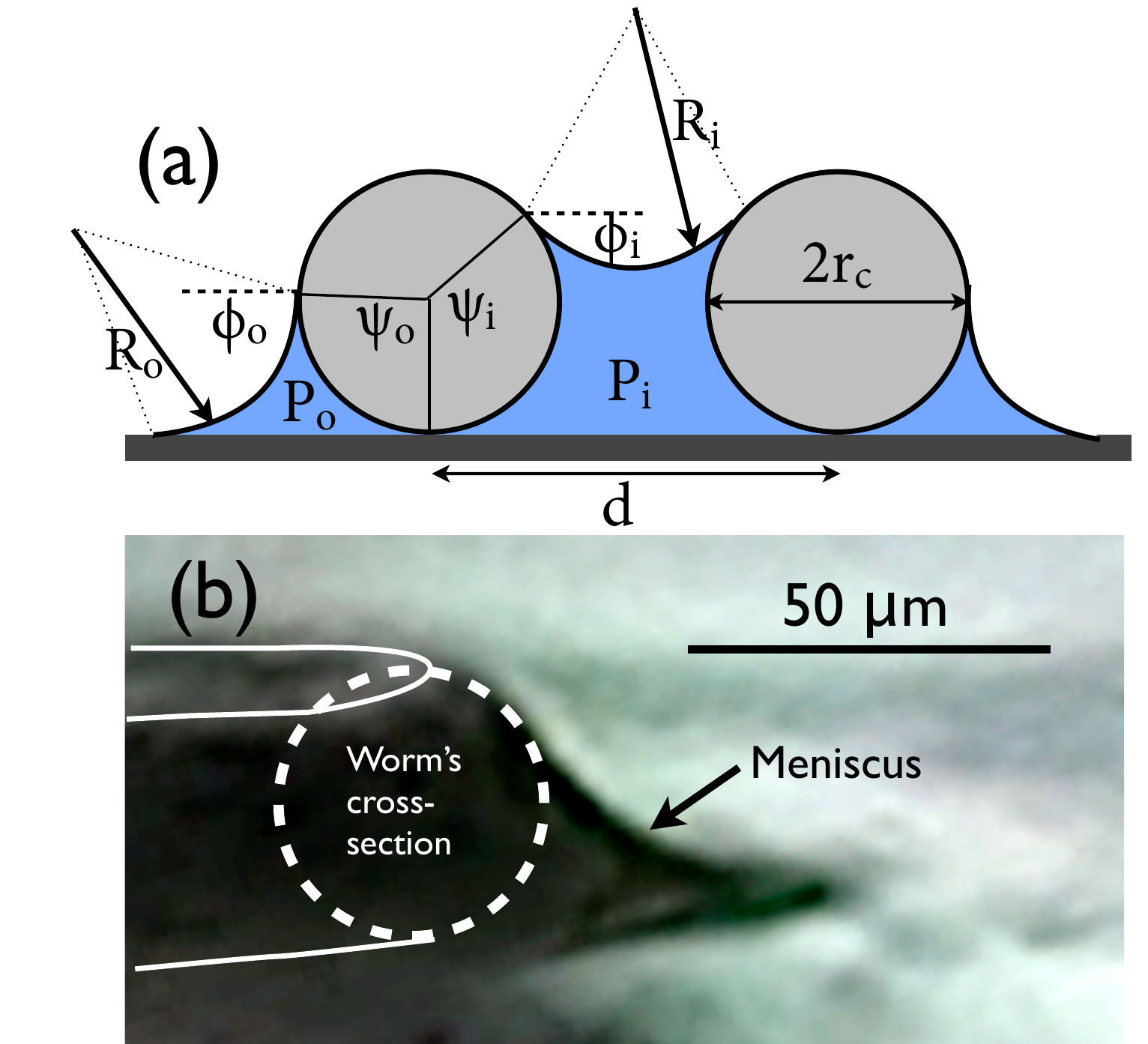}
\caption{(a) Schematic of cross-sectional view of two worms. (b) Experimental observations of meniscus around a moving nematode. (Supplementary Video 2$\ddag$)}
    \label{Fig_Schematic_calculation}
\end{figure}

To estimate the typical interaction force between two nematodes we calculate the typical force that `immersion forces' can exert on two horizontal cylinders. This has not been studied before although the force between floating cylinders has been calculated \cite{CHW1981,VM2005} as has the force between immersed spheres \cite{KN2001}. (The case of immersed objects is different from that of floating objects because in the former case the substrate provides whatever vertical force is required to satisfy the vertical force balance condition. The vertical force balance must be considered separately for floating objects.) We consider the idealised setup illustrated in Fig.~\ref{Fig_Schematic_calculation} (a), which is qualitatively similar to that observed for an actively moving nematode (see Fig.~\ref{Fig_Schematic_calculation} (b)).%The reason for this is presumably because the final result depends on the volume of liquid that is trapped between the objects. 

Both between the cylinders and outside, the angle of the meniscus to the horizontal satisfies $\phi = \pi - (\theta_c + \psi)$ at the contact line,
where $\theta_c$ is the contact angle on the cylinder. Elementary geometry then shows that the radii of curvature of the menisci in the inside and outsideregions are  
\beqs
R_i &=& \f{d-2 r_c \sin \psi_i }{2 \sin (\theta_c + \psi_i)} \,, \\ \nonumber
R_o &=& \f{r_c (1-\cos \psi_o)}{\cos\theta_g+\cos\bigl(\theta_c+\psi_o\bigr)} \,.
\eeqs
The total horizontal force consists of two components: a force due to surface tension acting at the contact line and a pressure force acting along the wetted perimeter. 
The horizontal force per unit length due to surface tension acting at the contact line is given by 
\beq
F_{\gamma}=\gamma (\cos \phi_i - \cos \phi_o) \,. 
\eeq
Here, a positive force corresponds to an attractive interaction.
There is also a contribution to the force from the difference in pressures between the inner and outer menisci. 
The pressure in the fluid in these regions is given by $ p-p_a = - \f{\gamma}{R}$.
The inward pressure force per unit length is then 
\beq
F_{p} =  \gamma r_c \left[ \int_{0}^{\psi_i}  \f{\sin \varphi}{R_i} ~\upd\varphi - \int_{0}^{\psi_o}\f{\sin \varphi}{R_o} ~\upd\varphi \right] \,.
\eeq
Therefore, the total horizontal force per unit length on the cylinder is
\beq
\f{F_{tot}}{\gamma} = \cos \phi_i - \cos \phi_o+\f{r_c}{R_i}(1-\cos\psi_i)-\f{r_c}{R_o}(1-\cos\psi_o) .
\label{totforce}
\eeq

In eqn.~\eqref{totforce} we have the force for given values of $\psi_i$ and $\psi_o$. 
However, we need to estimate these angles to estimate the total horizontal force.
To obtain an equation relating $\psi_o$ and $\psi_i$ we shall assume equal capillary pressures in the inside and outside regions, since then $R_i = R_o$. This may be justified physically since it is unlikely that nematodes will be in contact with the solid substrate all along their length. Hence any difference in capillary pressure would be quickly equalized.  We shall also make the simplifying assumption that the nematodes and the glass that they sit on are both perfectly wetting so that $\theta_c=0$.

\begin{figure}[ht] %% Exp_scheme 
    \centering
         \includegraphics[width=.4\textwidth]{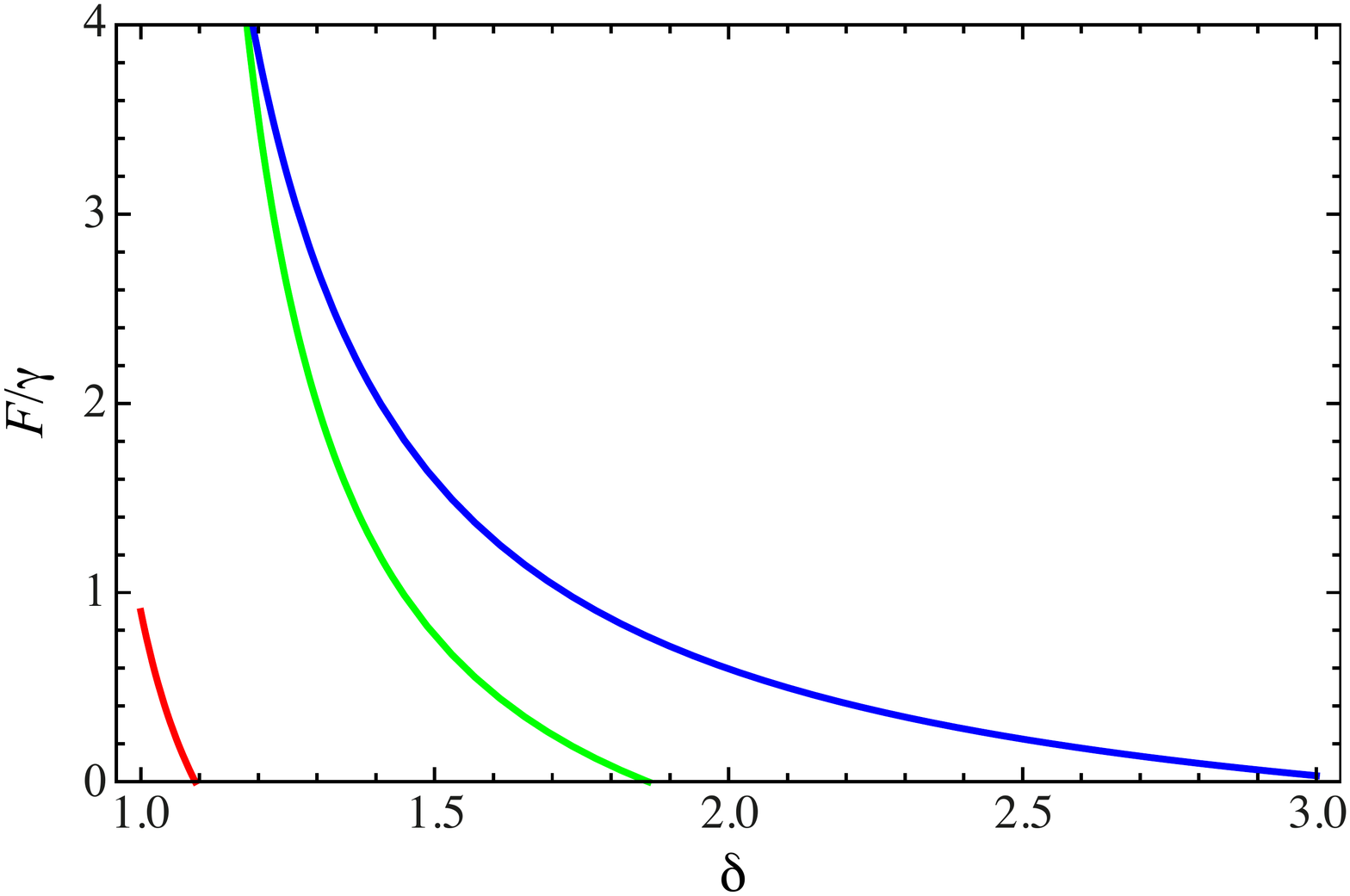}
\caption{ Force-displacement curves for different values of $\psi_i$: $\psi_i=1$ (red), $\psi_i=1.5$ (green) and $\psi_i=2$ (blue). Here $\delta=d/2r_c$ and we have assumed that $\theta=0$, as is likely to be the case for the worms.}
    \label{force_disp}
\end{figure}

With these simplifications, we find that the nonlinear terms cancel and we have simply
\beq
\cos\psi_o=\f{2}{\delta}\sin\psi_i-1
\eeq where $\delta=d/2r_c>1$. 
Therefore, the total horizontal force per unit length simplifies to
\beq
\f{F_{tot}}{\gamma}= \left[\f{2}{\delta}\sin\psi_i-\cos\psi_i-1\right]\left(1+\f{r_c}{R_i}\right) \,. 
\eeq 
For this force to be attractive, we require that
$ \f{2}{\delta}\sin\psi_i -\cos\psi_i>1$, that is, 
$\sin\bigl(\psi_i-\alpha\bigr)>\f{\delta}{\sqrt{\delta^2+4}}$ 
where 
$\tan\alpha=\delta/2 $.
For a given displacement, i.e.~given $\delta$, we have that the force is attractive provided that
\beq
\psi_i>\arcsin\left(\f{\delta}{\sqrt{\delta^2+4}}\right)+\arctan\f{\delta}{2}
\eeq

For simplicity, we shall assume that the contact line is pinned so that $\psi_i$ is constant as the cylinders move closer together. By plotting the force-displacement curves we find that, as expected, the attractive force increases as the two worms get closer together. Some typical results are shown in Fig. \ref{force_disp}. These plots also demonstrate that the force only acts over a lengthscale comparable to the radius of the worms; the force is relatively short ranged. This is consistent with the observation that zipping is driven by collision, rather than being driven by surface tension. Finally, the total horizontal force on a worm of length $1\mathrm{~mm}$ could easily reach the order of $50\mathrm{~\mu N}$, and consequently overwhelm the typical muscle force that can be exerted by a single nematode $\sim O(1)\mathrm{~\mu N}$.

\section{Discussion} 

We have studied the collective behaviour of nematodes in thin liquid layers, thereby uncovering the possibility of interactions between locomoting organisms without hydrodynamic effects. This effect is observed even when the nematodes are of very different sizes. We have shown that this interaction is not initially driven by surface tension forces but that once an interaction happens surface tension forces ensure that the nematodes remain aggregated. However, we have also shown that the characteristics of locomotion (namely the Strouhal number and body form) do not differ between individual and grouped nematodes. This result indicates that the body form is not significantly changed by the capillary force and, further, that no mechanical advantage is achieved. Nevertheless, grouped nematodes are observed to stay together for a long time (more than a few minutes in some cases). It is known that many other organisms move and live in circumstances where capillary forces are important and so it seems likely that, in some of these instances at least, the capillary force may be exploited to the organism's benefit, as has already been shown for individual water treaders, \emph{Mesovelia}\cite{HB2005}.
 
\section{Acknowledgements}

D.V.~is supported by an Oppenheimer Early Career Fellowship.

\end{document}